\documentclass[pdflatex,sn-mathphys-num]{sn-jnl}
\pdfoutput=1

\usepackage{graphicx}
\usepackage{amsmath,amssymb,amsfonts}
\usepackage{amsthm}
\usepackage{mathrsfs}
\usepackage[title]{appendix}
\usepackage{xcolor}
\usepackage{textcomp}
\usepackage{manyfoot}
\usepackage{booktabs}
\usepackage{algorithm}
\usepackage{algorithmicx}
\usepackage{algpseudocode}
\usepackage{listings}
\usepackage{subcaption}
\usepackage{caption} 
\makeatletter
\newcommand{\ExtendedDataFigures}{%
  \setcounter{figure}{0}%
  \renewcommand{\fnum@figure}{\textbf{Extended Data Fig.~\thefigure}~$|$}%
}

\long\def\@makecaption#1#2{%
  \vskip\abovecaptionskip
  \small
  #1 #2\par
  \vskip\belowcaptionskip
}
\makeatother

\begin{document}

\title{Tracking the Catastrophic Collapse of Hybrid Exciton-Phonon Order in a Quantum Material}

\author[1]{\fnm{Omar} \sur{Abdul-Aziz}} 
\author[1]{\fnm{Danilo} \sur{Comini}}
\author[2]{\fnm{Johannes} \sur{Lang}}
\author[2]{\fnm{Nils} \sur{Bartel}}
\author[3]{\fnm{Michael} \sur{Buchhold}}
\author[2]{\fnm{Sebastian} \sur{Diehl}}
\author[4]{\fnm{Daniel} \sur{Wolverson}}
\author[5]{\fnm{Charles J.} \sur{Sayers}}
\author[5,6]{\fnm{Giulio} \sur{Cerullo}}
\author*[1]{\fnm{Paul H. M.} \sur{van Loosdrecht}}
\email{pvl@ph2.uni-koeln.de}
\author*[1]{\fnm{Hamoon} \sur{Hedayat}}
\email{hedayat@ph2.uni-koeln.de}

\affil[1]{\orgdiv{II. Physikalisches Institut}, \orgname{Universit\"at zu K\"oln}, \orgaddress{\street{Z\"ulpicher Stra\ss e 77}, \city{K\"oln}, \postcode{D-50937}, \country{Germany}}}
\affil[2]{\orgdiv{Institut f\"ur Theoretische Physik}, \orgname{Universit\"at zu K\"oln}, \orgaddress{\street{Z\"ulpicher Stra\ss e 77}, \city{K\"oln}, \postcode{D-50937}, \country{Germany}}}

\affil[3]{\orgdiv{Institut für Theoretische Physik}, \orgname{Universität Innsbruck}, \orgaddress{\street{Technikerstraße 21a}, \city{Innsbruck}, \postcode{6020}, \country{Austria}}}
\affil[4]{\orgdiv{Department of Physics}, \orgname{University of Bath}, \orgaddress{\city{Bath}, \postcode{BA2 7AY}, \country{UK}}}

\affil[5]{\orgdiv{Dipartimento di Fisica}, \orgname{Politecnico di Milano}, \orgaddress{\street{Piazza L. da Vinci 32}, \city{Milan}, \postcode{20133}, \country{Italy}}}
\affil[6]{\orgdiv{Istituto di Fotonica e Nanotecnologie}, \orgname{Consiglio Nazionale delle Ricerche}, \orgaddress{\street{Piazza L. da Vinci 32}, \city{Milan}, \postcode{20133}, \country{Italy}}}
\maketitle

\begin{abstract}
    \p Revealing the hidden interactions that bind the electronic and lattice components of a cooperative quantum order is central to sculpting new states of matter. This challenge is epitomized by the charge density wave material 1\textit{T}-TiSe$_2$, where photoexcitation disrupts its presumed hybrid exciton-phonon order, exposing a striking paradox: the electronic component collapses within femtoseconds, while the periodic lattice distortion persists~\cite{porerNonthermalSeparationElectronic2014}, challenging the very definition of a hybrid order: if the lattice distortion outlives the excitonic condensate, were they ever truly intertwined? Here we resolve this paradox by uncovering a low-frequency mode ($\sim 0.13~\text{THz}$) that emerges only in the ordered state and signals the presence of exciton-phonon coupling. This mode is consistent with a locked phason, a collective excitation that would arise if the coupling between the excitonic condensate and the lattice reduced its continuous phase symmetry to a discrete one, thereby giving the excitonic Goldstone mode a finite mass. Such a scenario is captured by an effective theory, which describes a shared potential landscape linking the excitonic and lattice degrees of freedom. At a critical photoexcitation threshold, the collapse of the excitonic order flattens the potential, triggering an exciton-phonon catastrophe characterized by selective overheating of the charge density wave phonon, the disappearance of the locked phason, and a sudden loss of electronic coherence. Remarkably, the lattice distortion survives this event as a dynamically trapped and non-thermal remnant, whose non-equilibrium character is confirmed by the anomalous temperature dependence of the phononic response. These findings demonstrate that the coupled potential energy landscape of cooperative orders can be manipulated to selectively dismantle complex quantum orders, advancing a new paradigm for material control through dynamical design.
\end{abstract}

\keywords{Charge Density Waves, Ultrafast Dynamics, Time-Resolved Raman Spectroscopy, Excitonic Insulator, Coupled Order Parameters, 1\textit{T}-TiSe$_2$}

\section*{Introduction}

In the landscape of quantum materials, the emergence of cooperative phases from the delicate interplay of electronic and lattice degrees of freedom is a central theme. Phenomena like charge density wave (CDW), unconventional superconductivity, and Mott insulating states are manifestations of this intricate coupling, where feedback between electron-phonon interactions and correlation-driven instabilities dictates the ground state~\cite{rowe2023resonant,maklarNonequilibrium2021}. A key frontier is therefore to transcend the static description of these phases, which, while well-defined, often conceals the complex hierarchy of interactions responsible for the phase stability. By leveraging intense ultrafast light-matter interactions, it is possible to break this equilibrium balance and drive a material into a non-thermal regime. The ultimate goal is to sculpt and control novel light-driven states of matter, transient configurations with emergent functionalities beyond the equilibrium limits, and establish a new paradigm of materials through dynamic design~\cite{yusupov2010coherent,domroseLightinduced2023}.

Quantum materials with coupled electronic and lattice orders provide a fertile ground for exploring complex many-body phenomena, where the hierarchy of interactions challenges conventional understanding and motivates advances in both theory and experiment, enabling the discovery of emergent phases under controlled perturbations. The layered transition metal dichalcogenide 1\textit{T}-TiSe$_2$ has long served as the prototypical testbed for this pursuit. Below a transition temperature of $T_\mathrm{CDW} \approx 202$~K, it hosts a commensurate CDW that simultaneously features a 2$\times$2$\times$2 periodic lattice distortion (PLD) and the opening of an indirect bandgap~\cite{traum1978ti}. The fundamental origin of this phase has been the subject of a long and vigorous debate, with initial theories proposing a phonon-driven instability~\cite{williams1976phase,hughes1977structural,rossnagel2002charge,weberElectronPhononCouplingSoft2011} or pointing towards a purely excitonic insulator mechanism~\cite{di1976electronic,pillo2000photoemission,cercellier2007evidence,monneyTemperaturedependent2010,rohwerCollapse2011,rossnagel2011origin}. This dichotomy has now largely been resolved into the prevailing consensus that the CDW is a cooperative phenomenon~\cite{van2010exciton,porerNonthermalSeparationElectronic2014}. In the equilibrium ground state, the excitonic and structural instabilities are not competing; rather, they are mutually reinforcing, locking the system into a hybrid order where the two components are inseparable. Yet, while this cooperative picture is widely accepted~\cite{kurtzNonthermal2024,novkoElectronCorrelationsRule2022,kanekoExcitonphononCooperativeMechanism2018,ottoMechanisms2021,michaelOptical2022}, a direct experimental signature of the coupling itself has remained elusive.

To resolve this, it is essential to identify a spectroscopic observable that captures the coupling between the electronic condensate and the lattice. It is well established that the electronic gap oscillates at the frequency of the CDW phonon ($A_{1g}^*$)~\cite{monneyRevealing2016,hedayatExcitonicLatticeContributions2019,duanOptical2021}, yet this behavior alone does not constitute proof of a cooperative mechanism; such oscillations can also arise in a Peierls‑like CDW, where the electronic order parameter follows the lattice distortion dynamics~\cite{hellmannTime2012}. Following the experimental observation of exciton condensation in this system~\cite{kogarSignatures2017}, the key question now is whether the condensate exhibits its own intrinsic collective dynamics that actively couple to the lattice, thereby revealing the true cooperative nature of the order. The excitonic insulator phase is described by a complex order parameter $\Delta = |\Delta| e^{i\varphi}$, whose amplitude and phase reflect the magnitude and coherence of the electron-hole condensate. In the presence of a continuous U(1) symmetry, fluctuations of the phase $\varphi$ cost negligible energy and correspond to a massless collective mode. In this context, the U(1) symmetry corresponds to the global phase of the electron-hole condensate, and the coupling to the CDW phonon explicitly breaks this continuous symmetry by fixing the relative phase between electronic and lattice components, thereby generating a pseudo-Goldstone mass for the phason, i.e., the locked phason (Fig.~\ref{fig:cdw_raman_overview}a)~\cite{murakami2020collective}. The phason frequency $\omega_\mathrm{phason}$ thus serves as a direct spectroscopic fingerprint of the coupled exciton-phonon condensate and its symmetry-broken ground state. In CDWs driven by strong electron-phonon coupling, commensurability and strong bonding forces impose a rigid potential, producing a stiff locked phason with frequency comparable to that of the CDW phonon $\Omega$, in the $ \mathrm{THz}$ range~\cite{sugai2006phason,baldinispontaneous2023}. By contrast, in a hybrid excitonic regime, the mass arises from comparatively weak exciton-phonon coupling, yielding a soft locked phason with $\omega_\mathrm{phason} \ll \Omega$~\cite{ningSignatures2020,mazza2020nature,zenker2014fate}. Although low-frequency locked phasons have been observed in incommensurate CDW where continuous symmetry degeneracy is lifted by interactions such as impurities, interlayer coupling, or Coulomb forces
~\cite{shengTerahertz2024,kimObservation2023,baeDynamic2025}, their detection in a commensurate CDW system such as 1\textit{T}-TiSe$_2$  supports an excitonic-phononic hybrid origin~\cite{van2010exciton,porerNonthermalSeparationElectronic2014,kanekoExcitonphononCooperativeMechanism2018}. This distinction renders the locked phason a spectroscopic fingerprint of cooperative exciton-phonon coupling.

The equilibrium picture of a unified, inseparable hybrid order is challenged when the system is driven far from equilibrium. A wealth of time-domain studies on 1\textit{T}-TiSe$_2$ have revealed a striking paradox~\cite{porerNonthermalSeparationElectronic2014,hedayatExcitonic2019,burianStructural2021,chengLightinduced2022,duanOpticalManipulationElectronic2021,ouIncoherencetocoherenceCrossoverObserved2024,lianUltrafast2020,chengUltrafastFormationTopological2024,kurtzNonthermal2024}: Probes sensitive to electronic order, particularly time- and angle-resolved photoemission spectroscopy (TR-ARPES),  report a collapse of the CDW gap within 100~femtoseconds ~\cite{rohwerCollapse2011} after photoexcitation. In stark contrast, ultrafast diffraction experiments reveal that the superlattice peaks associated with the PLD persist for several picoseconds. Recent ultrafast low-energy electron diffraction studies interpreted the persistence of the PLD as evidence for a residual Peierls component of the CDW~\cite{kurtzNonthermal2024}. This subtlety suggests that both the excitonic and lattice components contribute but ultimately remain separable. In this view, once the excitonic condensate is disrupted by photoexcitation, a structural distortion remains, attributed to an underlying lattice-driven instability that survives independently. However, such an interpretation treats the excitonic and structural components as cooperative yet fundamentally distinct, challenging the notion of a truly hybrid order parameter and raising deeper questions about the physical origin and nature of the surviving PLD.  While several observations of this transient decoupling have been made, the mechanism governing this phenomenon remains unresolved. Time-resolved Raman spectroscopy (TR-Raman) is ideally suited to address this, as it directly probes the dynamics of CDW phonons as well as the electronic susceptibility. Despite its proven sensitivity to non-equilibrium collective dynamics in correlated materials~\cite{yangUltrafast2020,chouUltrafast2024,glierDirectObservationHiggs2024}, it has yet to be fully exploited to unravel the dynamics of CDW systems. 

In this work, we resolve the paradox of hybrid order in 1\textit{T}-TiSe$_2$ by identifying a low-energy ($\sim$0.13~THz) collective mode that emerges exclusively in the ordered phase. Its appearance provides a spectroscopic signature of coupling between the excitonic and lattice subsystems. The frequency and non-equilibrium response of this mode are consistent with the behavior expected for a locked phason, arising from the coupling between the excitonic condensate and the CDW phonon. At a critical photoexcitation threshold, the collapse of the excitonic order parameter $\Delta$ flattens the coupled potential landscape, suppressing the restoring force for the CDW phonon and triggering a cascade of effects linked to an exciton-phonon catastrophe: phonon overheating, locked phason suppression, and the emergence of a non-thermal metastable state. The resulting phase, characterized by a saturated phonon response and the absence of the low-energy hybrid phason, is a dynamically trapped, non-thermal remnant of PLD, whose nature is confirmed by a temperature evolution of the phonon component that diverges from the equilibrium CDW Raman response, indicating a distinct non-equilibrium trajectory. These findings establish a framework for disentangling and controlling cooperative quantum orders through targeted instabilities of a coupled potential landscape.

\subsection*{Raman Response of the CDW phonons in 1\textit{T}-TiSe$_2$}
\begin{figure}[ht!]
\centering
\includegraphics[width=\linewidth]{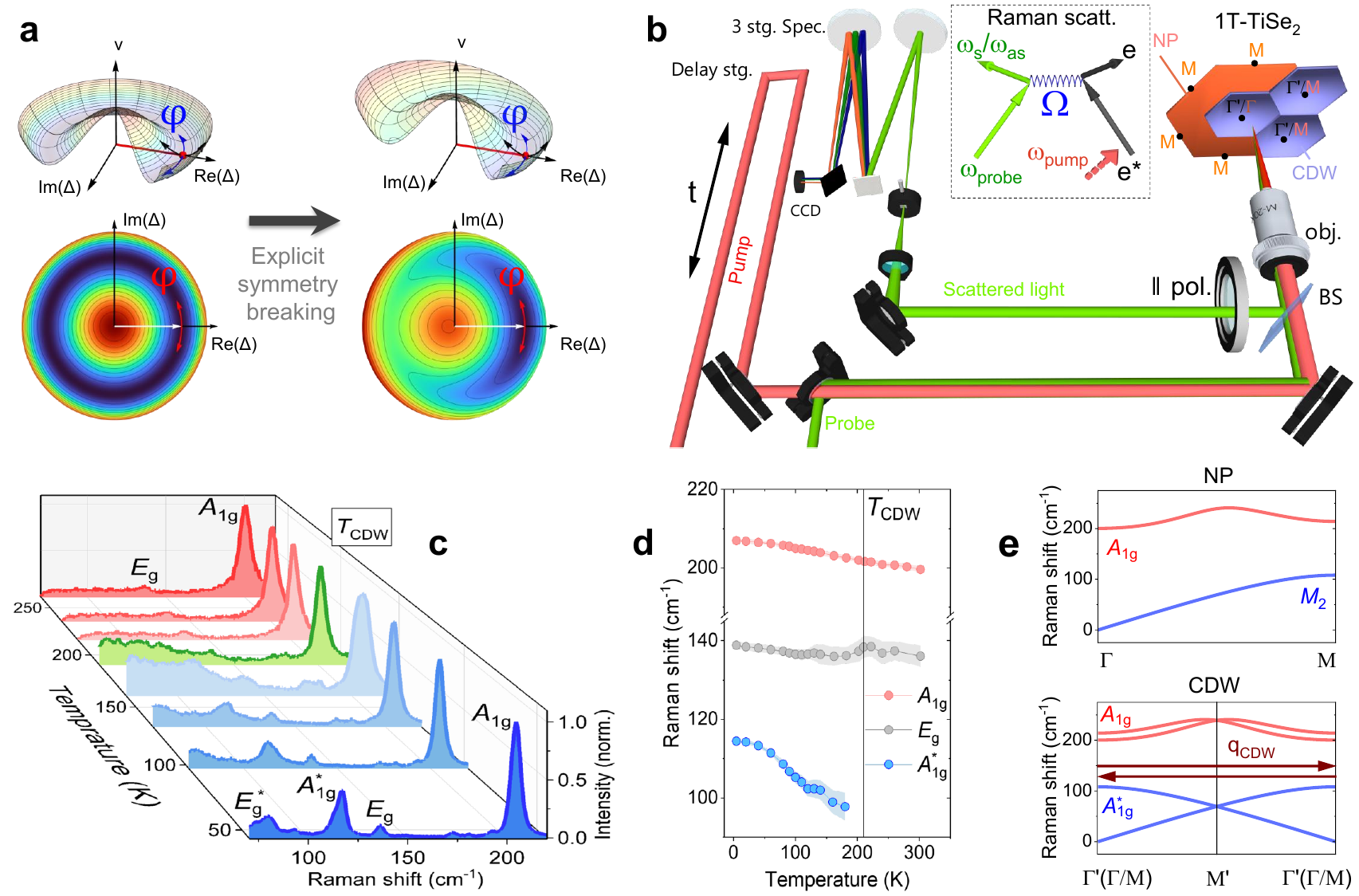}
\caption{\textbf{Raman access to coupled order in 1\textit{T}-TiSe$_2$.} 
\textbf{a,} Schematic free-energy landscape of the complex order parameter $\Delta = |\Delta| e^{i\varphi}$. Left: in the absence of locking, the potential has a Mexican-hat shape with a degenerate circle of minima in the $(\mathrm{Re}\,(\Delta), \mathrm{Im}\,(\Delta))$ plane, and the phason corresponds to motion along the phase coordinate $\varphi$. Right: pinning to an external potential, e.g., exciton-phonon interactions, lifts this degeneracy, thereby turning the phason into a locked finite-frequency mode. 
\textbf{b,} Overview of the experimental configuration and   processes. Time-resolved Raman spectroscopy setup combining a femtosecond 750~nm pump with a time-delayed 513~nm probe in parallel backscattering geometry. Inset: Pump-probe Raman process; the pump pulse excites electrons, which relax by emitting phonons; the delayed probe detects Stokes (phonon creation) and anti-Stokes (phonon annihilation) photons. 
\textbf{c,} Temperature-dependent Raman spectra under weak CW excitation at 532~nm. Below $T_\mathrm{CDW} \approx 202$~K, folded $A_{1g}^*$ and $E_g^*$ modes appear.
\textbf{d,} Evolution of phonon energies. The $A_{1g}^*$ mode hardens upon cooling, with $A_{1g}^*$ near 116~cm$^{-1}$ at low temperature, serving as the primary phononic marker of the CDW. 
\textbf{e,} Schematic phonon dispersions from DFT in the normal phase (NP, top) and in the CDW phase (CDW, bottom), showing only the branches relevant for the discussion. In the NP, the $A_{\mathrm{1g}}$ branch (red) is Raman active at $\Gamma$, while the soft $M_2$ branch (blue) resides at the zone-boundary M point and is therefore not accessible in first-order Raman scattering. In the CDW phase, Brillouin-zone folding maps M onto the new zone centre $\Gamma'$, where the former $M_2$ branch appears as the folded $A_{\mathrm{1g}}^{\mathrm{*}}$ mode (blue). The emergence of $A_{\mathrm{1g}}^{\mathrm{*}}$ at $\Gamma'$ in the Raman spectra is thus a direct fingerprint of the PLD.}
\label{fig:cdw_raman_overview}
\end{figure}
Figure~\ref{fig:cdw_raman_overview}b illustrates the TR-Raman experiment: a femtosecond 750~nm pump pulse photoexcites charge carriers, while a $\sim$1.2~ps and 513~nm probe pulse captures both Stokes and anti-Stokes Raman scattering in a parallel polarization scheme using a triple-stage spectrometer and CCD detector (see Sec.~Methods and Supplementary Note~3). The inset illustrates the Raman process: the ultrafast photoexcitation promotes electrons across the indirect gap, creating a non-equilibrium carrier population that couples to a phonon of energy $\Omega$. The time-delayed probe then drives a virtual electronic transition and is inelastically scattered as a Stokes ($\omega_{0} - \Omega$) or anti-Stokes ($\omega_{0} + \Omega$) photon, thus reading out the phonon population.

To establish a baseline for the coupled electron-lattice dynamics, we first map the equilibrium Raman spectrum of 1\textit{T}-TiSe$_2$ across its CDW transition.  Figure~\ref{fig:cdw_raman_overview}c displays temperature‑dependent Raman spectra obtained under weak continuous‑wave excitation in a parallel polarization configuration. Above $T_\mathrm{CDW}$, only the zone‑center $A_{\mathrm{1g}}$ and $E_{\mathrm{g}}$ modes of the high‑symmetry phase are visible. Cooling below the transition activates two additional peaks, $A_{\mathrm{1g}}^{\mathrm{*}}$ and $E_{\mathrm{g}}^{\mathrm{*}}$ of the CDW phase.

\begin{figure}[p]
\centering
\includegraphics[width=1\textwidth]{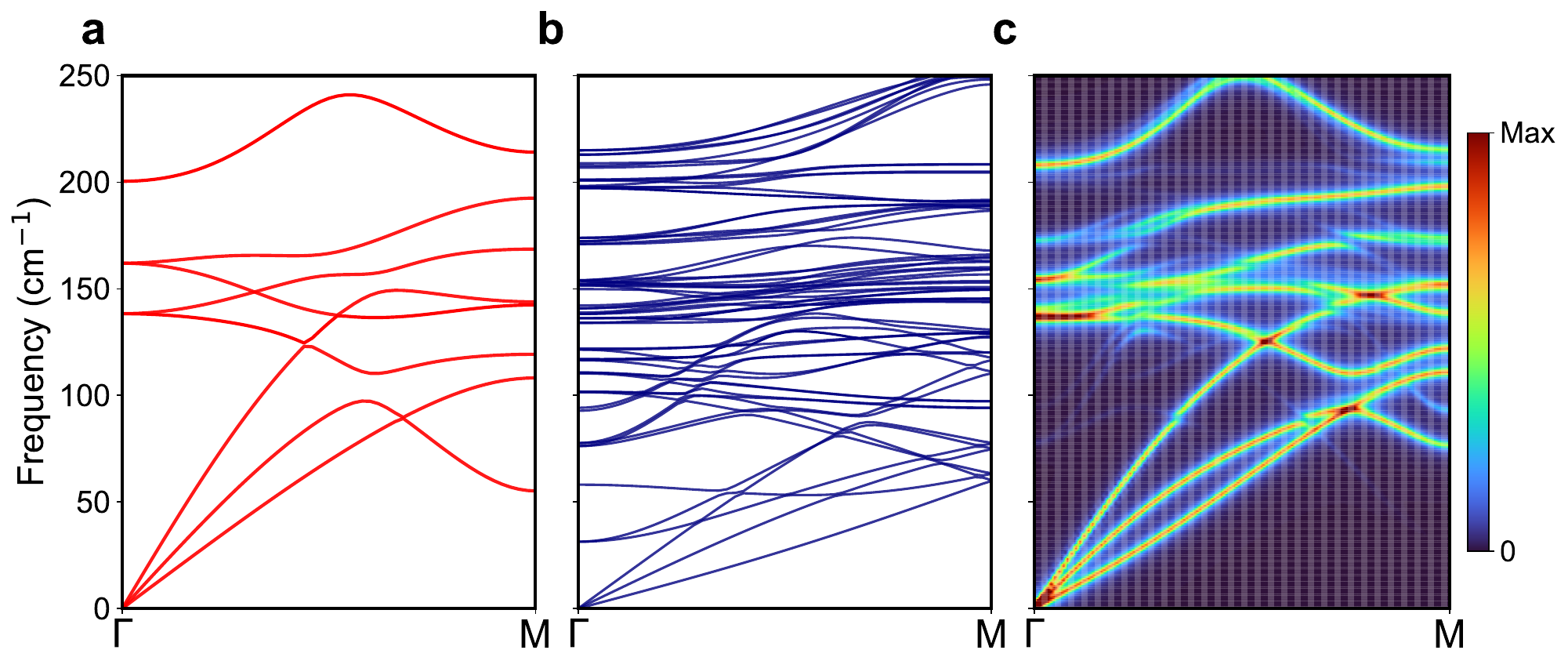}
\caption{\textbf{Phonon dispersion of  1\texorpdfstring{$T$}{T}-TiSe\(_2\).} 
(a) Phonon dispersion in the normal phase (space group \(P\bar{3}m1\)) from DFT simulations. 
(b) Phonon dispersion in the CDW phase (space group \(P\bar{3}c1\)). The \(A_{1g}^{*}\) and \(E_{g}^{*}\) modes originate from M (or L)-point phonons that fold to the new \(\Gamma'\) point in the reduced Brillouin zone, consistent with the zone-folding picture of the CDW. 
(c) Unfolded phonon dispersion of the CDW phase of TiSe\(_2\). The false-colour scale (arbitrary units), indicated by the colour bar on the right, shows the weights of the projections of the CDW phonon branches onto the irreducible representations of the little groups of the 3-atom primitive structure. The \(k\)-point labels refer to the unfolded (primitive) Brillouin zone, from \(\Gamma\) to \(M\).  The phonon unfolding was performed using the UPHO code \cite{PhysRevB.95.024305}.
}
\label{ED1}
\end{figure}

In TR-Raman, to track the evolution of the CDW, we focus on the $A_{\mathrm{1g}}^{\mathrm{*}}$ phonon. The temperature dependence of the static Raman spectra (Fig.~\ref{fig:cdw_raman_overview}d) reveals that this mode emerges just below $T_\mathrm{CDW}$ at $\sim$95~cm$^{-1}$ and hardens progressively to $\sim$116~cm$^{-1}$ upon cooling, whereas the parent-phase $A_{\mathrm{1g}}$ mode remains only slightly affected~\cite{duongRamanCharacterizationCharge2017}. First-principles phonon calculations (Fig.~\ref{fig:cdw_raman_overview}e and Extended Data Fig.~\ref{ED1}) confirm that the $A_{\mathrm{1g}}^{\mathrm{*}}$ mode originates from the $M$-point phonon (labelled $M_2$) of the normal phase, identified as the CDW soft mode~\cite{weberElectronPhononCouplingSoft2011,kogarSignatures2017}, which folds to the Brillouin-zone center $\Gamma'$ of the $2\times2\times2$ superstructure (Fig.~\ref{fig:cdw_raman_overview}a,e). Since Raman spectroscopy probes only zone-center excitations, the appearance of this folded mode provides direct spectroscopic evidence of symmetry breaking and the formation of the PLD~\cite{holyRaman1977}.  Crucially, the $A_{\mathrm{1g}}^{\mathrm{*}}$ phonon is not merely a marker of lattice distortion in Raman response. This CDW phonon strongly modulates the valence-band conduction-band hybridization amplitude~\cite{monneyRevealing2016} and, by sharing the same wavevector $q_{\mathrm{CDW}}$, couples linearly to the excitonic condensate~\cite{kogarSignatures2017}. This interaction renders it a highly sensitive probe of the hybrid order in 1\textit{T}-TiSe$_2$, governed by the cooperative electronic and lattice dynamics.

\subsection*{Critical transformation of the CDW phonon under photoexcitation}

\begin{figure}[ht!]
\centering
\includegraphics[width=\linewidth]{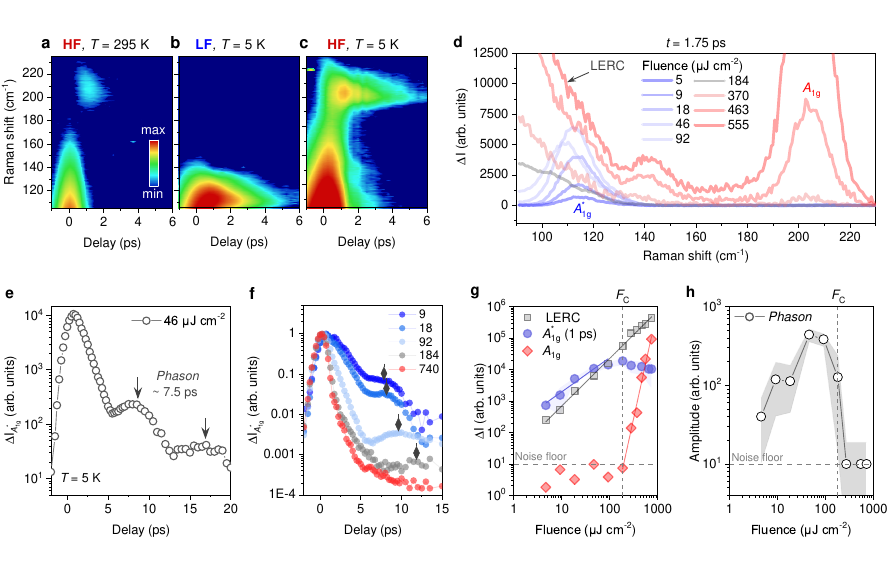}
\caption{\textbf{Critical transformation of the CDW phonon under photoexcitation.}
\textbf{a--c,} Time-resolved Raman maps of the pump-induced differential intensity $\Delta I(t,\omega)$ under three representative conditions. 
\textbf{a,} High fluence at 295~K: $\Delta I$ shows a strong low-energy Raman continuum (LERC) and population of the normal-phase $A_{\mathrm{1g}}$ near 200~cm$^{-1}$. 
\textbf{b,} Low fluence at 5~K: $\Delta I$ contains LERC and the folded CDW phonon $\Delta I_{A_{1g}^{*}}$. 
\textbf{c,} High fluence at 5~K: the LERC and $\Delta I_{A_{1g}}$ dominate the spectrum, and $\Delta I_{A_{1g}^{*}}$ becomes spectrally obscured by the LERC. 
\textbf{d,} Differential spectra $\Delta I(\omega)$ at a fixed delay of 1.75~ps for fluences as indicated. $\Delta I_{A_{1g}^{*}}$ (blue) dominates at low fluence, weakens near the threshold $F_c \approx 187~\mu$J\,cm$^{-2}$, and the normal-phase $\Delta I_{A_{1g}}$ (red) emerges abruptly above threshold, while the broadband $\Delta I_{\mathrm{LERC}}$ increases steadily. 
\textbf{e,} Time trace of the CDW mode $\Delta I_{A_{1g}^{*}}$ at 46~$\mu$J\,cm$^{-2}$ and 5 K, showing that the slowly slowly decaying dynamics are modulated by a $\sim$7.5~ps (0.13~THz) oscillation (arrows), revealing the locked phason coupled to $A_{\mathrm{1g}}^{\mathrm{*}}$. 
\textbf{f,} Dynamics of $A_{\mathrm{1g}}^{\mathrm{*}}$ for fluences as indicated, the oscillation amplitude decreases while its period increases, indicating locked phason softening (see diamonds). 
\textbf{g,} Fluence dependence of the extracted intensities: $\Delta I_{\mathrm{LERC}}$ grows superlinearly ($\propto F^{1.51\pm0.08}$); $\Delta I_{A_{1g}}$ appears only above $F_c$ and shows extreme nonlinearity ($\propto F^{6.99\pm0.40}$); the amplitude of $\Delta I_{A_{1g}^{*}}$ initially scales linearly with fluence ($\propto F^{1.00 \pm 0.10}$), but saturates as $F \geq F_c$. 
\textbf{h,} The locked phason amplitude grows only within a narrow low-fluence range and vanishes near $F_c$, revealing a distinct behavior that identifies it as a separate collective excitation.}
\label{fig:2}
\end{figure}

The TR-Raman response of 1\textit{T}-TiSe$_2$ reveals that the hybrid order is highly susceptible to ultrafast perturbations, undergoing a qualitative transformation once the photoexcitation density exceeds a well-defined threshold. At room temperature ($T\!\gg\!T_\mathrm{CDW}$), the sample is in the undistorted normal phase. Fig.~\ref{fig:2}a shows that immediately after photoexcitation, a low-energy Raman continuum (LERC) at $ \lesssim 150$  \,cm$^{-1}$ rises and decays within the $\sim$1.2\,ps pump-probe cross-correlation, followed by a transient population of the normal-phase mode near $200$\,cm$^{-1}$, indicated by the change in its Raman intensity $\Delta I_{A_{1g}}$. The rise of the LERC mirrors the instrumental response function and resembles the fast sub-picosecond component in the time-resolved reflectivity (TRR) traces (Supplementary Note~11), pointing to ultrafast electronic energy redistribution rather than phonon population dynamics. While broad continua are known in equilibrium Raman spectra of other excitonic materials, such as Ni$_2$TaSe$_5$~\cite{volkovCriticalChargeFluctuations2021,kimPhononicSoftMode2020}, we avoid assigning this feature as a direct signature of the exciton condensate here. Instead, given its nearly invariant lineshape across conditions, we regard it as an electronic background channel (i.e., a change in the Raman susceptibility), and model it using an exponential function at each delay, which is then subtracted to provide the isolated phonon response (Supplementary Note~4).

At $T\!=\!5$\,K, there is a clear difference in the non-equilibrium response. For low fluences (LF; $F\!\lesssim\!100~\mu$J\,cm$^{-2}$), the dominant feature is a photoinduced change in the CDW phonon intensity $\Delta I_{A_{1g}^{*}} $, while the normal-phase phonon intensity remains unaffected by the pump. The LERC increases only modestly and, after approximately 1.5\,ps, decays sufficiently to reveal a clean phononic spectrum.

In the high-fluence (HF) regime ($F\!\gtrsim\!200~\mu$J\,cm$^{-2}$; Fig.~\ref{fig:2}a, right), the $\Delta I_{A_{1g}}$ response becomes visible, with an intensity even greater than that observed at 300\,K, while the LERC intensifies and spectrally overwhelms $\Delta I_{A_{1g}^{*}} $, making the latter difficult to resolve. To quantify this crossover, we analyze the fluence dependence at a fixed delay of 1.75\,ps (Fig.~\ref{fig:2}b). As $F$ increases, the $\Delta I_{A_{1g}^{*}} $ intensity initially increases, but near $F\!\simeq\!187~\mu$J\,cm$^{-2}$ its spectral weight above $\sim\!100$\,cm$^{-1}$ begins to diminish. Concurrently, the $\Delta I_{A_{1g}}$ associated with the normal phase exhibits a photoinduced increase starting at $F\!\gtrsim\!200~\mu$J\,cm$^{-2}$. This transition becomes apparent only above a critical threshold around $F_c\!\approx\!100$-$200~\mu$J\,cm$^{-2}$. Based solely on this, one might speculate that the coexistence of an increasing $\Delta I_{A_{1g}}$ with a suppressed $\Delta I_{A_{1g}^{*}} $ reflects a photoinduced drift from the CDW landscape toward the normal phase; however, as shown below, this inference does not withstand scrutiny.

The most striking feature in the dynamics is a long-period modulation at $\sim$ 0.13\,THz which modulates the \(A_{1g}^{*} \) intensity (Fig.~\ref{fig:2}e), and is inconsistent with conventional phonon dynamics and points to a collective mode of the CDW with finite
frequency, enabled by the hybrid exciton-phonon order (see Fig.~\ref{fig:2}f). We systematically rule out alternative scenarios. (i) A purely phononic origin is excluded, as no low-energy branches are found in either the normal or CDW phase, based on existing literature~\cite{bianco2015electronic} and our DFT calculations. (ii) Beating between nearly degenerate phonons is inconsistent with both the fluence dependence of the oscillation and the absence of any such modes in our spectral window. (iii) A plasmonic origin is likewise implausible, as plasmon modes typically appear at much higher frequencies (\( \gtrsim 1 \)~THz)~\cite{porerNonthermalSeparationElectronic2014,kogarSignatures2017}. (iv) Phase stiffness governed by lattice pinning or impurity potentials fails to account for the observed behavior of the mode. In commensurate CDW systems, the energy scale of phason is set by the Peierls pinning potential, i.e., comparable to bare lattice frequencies in the THz range. However, if the mode were of electron-phonon-driven, Peierls-type origin, photoexcitation would harden the phase mode toward its bare lattice frequency~\cite{thomsonPhase2017}, in stark contrast to the softening observed here. Moreover, the abrupt disappearance of the 0.13\,THz mode at a fluence where the CDW phonon remains clearly visible further supports that it does not originate from lattice pinning and thus rules out a purely lattice-driven origin for the observed low-frequency oscillation.  (v) Finally, instrumental artifacts or modulation of the Raman tensor are excluded, as they cannot generate a discrete frequency component with robust phase coherence and strong fluence-dependent behavior (Supplementary Notes~5-7). 

Collectively, the evidence supports the interpretation of this mode as a locked phason of a hybrid exciton-phonon CDW. Several key observations reinforce this assignment. (i) The mode exclusively modulates the \( A_{1g}^{*} \) phonon and is absent in other phononic or electronic dynamics (Supplementary Note~7). (ii) Its normalized amplitude, the ratio of the oscillation to the underlying \( A_{1g}^{*} \) mode, remains constant for fluences \( F \lesssim 20~\mu\mathrm{J\,cm}^{-2} \), a regime in which the excitonic condensate remains largely intact~\cite{porerNonthermalSeparationElectronic2014,hedayatExcitonicLatticeContributions2019}, while the frequency of the oscillation softens continuously as the system approaches the critical threshold \( F_c \) (Fig.~\ref{fig:2}f, Extended Data Fig.~\ref{ED2}).  (iii) The mode vanishes abruptly at \( F_c \), even though the spectral weight and frequency of the \( A_{1g}^{*} \) phonon persist well beyond this threshold. Throughout this work, we define \( F_c \)  as the excitation fluence at which the locked phason amplitude collapses, marking the loss of coherent excitonic order while the lattice distortion remains.

Reports of locked phasons in other CDW systems~\cite{shengTerahertz2024,kimObservation2023} support the presence of collective modes in the 0.1-0.15\,THz range, but the fluence-dependent softening and abrupt disappearance observed here provide direct evidence of a hybrid character. We thus interpret the 0.13\,THz oscillation as a \textit{locked phason} of the coupled exciton-phonon order, i.e., a pseudo-Goldstone mode of the excitonic order which is pinned to the lattice. Its small gap arises from the breaking of the excitonic symmetry by phonon interactions, and its collapse at \( F_c \) reflects the loss of excitonic order, even as a metastable PLD remains. In the following sections, we demonstrate how such a locked phason emerges naturally in the effective theory of the exciton-phonon coupled potential.

To deconstruct the competing dynamics within the non-equilibrium state, we performed a quantitative analysis of the distinct components of the Raman response by fitting each spectrum with constrained Gaussians for the $A_{\mathrm{1g}}^{\mathrm{*}}$ and $A_{\mathrm{1g}}$ modes on top of an exponential LERC (for details of the analysis, Supplementary Note~4). The extracted fluence dependencies, shown in Fig.~\ref{fig:2}g, reveal that the electronic continuum, the normal-phase phonon, and the CDW phonon are governed by fundamentally different physical mechanisms.  First, the photoinduced change of LERC intensity grows superlinearly as $\Delta I_{\mathrm{LERC}}\!\propto\!F^{1.51\pm0.08}$ (Fig.~\ref{fig:2}e). This non-integer exponent signifies a complex electronic response beyond simple single-photon absorption, likely involving a combination of non-thermal carrier-carrier scattering and a strong modulation of the electronic Raman susceptibility. Next, the changes in the normal-phase $A_{\mathrm{1g}}$ mode appear only above the critical threshold, and the amplitude rises with an exceptionally large exponent, $\Delta I_{A_{1g}}\!\propto\!F^{6.99\pm0.40}$. Such extreme nonlinearity is the hallmark of a highly cooperative or avalanche-like activation channel, which we attribute to a multiphonon-assisted carrier relaxation pathway across the residual partial gap, a scenario consistent with TR-ARPES observations of surviving gap features above $F_c$~\cite{hedayatExcitonicLatticeContributions2019,duanOptical2021}. For \( F \lesssim 20~\mu\mathrm{J\,cm}^{-2} \), the change in the \( A_{1g}^{*} \) amplitude scales linearly with fluence,
\( \Delta I_{A_{1g}^{*}} \propto F^{1.00 \pm 0.10} \), consistent with a linear-response regime in which the CDW order is only weakly disturbed. Deviations from this behaviour appear only for \( F \)~$\geq$~\( F_c \), where the \( A_{1g}^{*} \) response saturates, signalling the onset of a distinct state in the hybrid exciton-phonon potential landscape (Fig.~\ref{fig:2}g). The persistence of this spectral weight, even in a regime dominated by the normal-phase mode, provides crucial evidence against a simple thermal melting of the CDW, which would require the CDW phonon to soften and disappear rather than saturate while retaining its spectral weight, supporting the survival of a PLD remnant in a non-equilibrium state.  Finally, the amplitude of the locked phason exhibits a qualitatively different fluence dependence from all other observables: it grows only within a narrow low-fluence window, and then collapses to zero as \( F \)~$\geq$~\( F_c \) (Fig.~\ref{fig:2}g). This distinct trend shows that the phason does not simply follow the $A_{\mathrm{1g}}^{\mathrm{*}}$ intensity or the electronic background, but instead represents a separate collective excitation.

\subsection*{Selective overheating and saturation of the CDW phonon}

\begin{figure}[ht!]
\centering
\includegraphics[width=\linewidth]{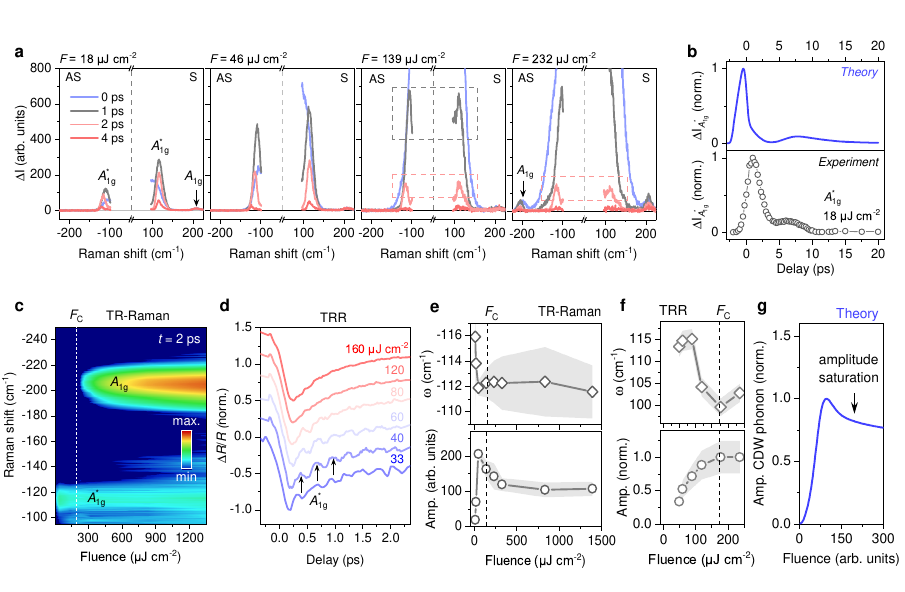}
\caption{\textbf{Selective overheating and saturation of the CDW phonon.}
\textbf{a,} Time-resolved anti-Stokes (AS) and Stokes (S) Raman spectra at selected delays for increasing fluences. At $F = 139~\mu$J\,cm$^{-2}$, the anti-Stokes and Stokes amplitudes of the $A_{\mathrm{1g}}^{\mathrm{*}}$ mode become nearly equal (dashed box), $I_\mathrm{AS} \approx I_\mathrm{S}$, indicating a strongly overheated, highly populated $A_{\mathrm{1g}}^{\mathrm{*}}$ phonon. Above $F_c$, the $A_{\mathrm{1g}}^{\mathrm{*}}$ response saturates and an AS-active contribution from the normal-phase $A_{\mathrm{1g}}$ mode emerges. 
\textbf{b,} Delay-dependent intensity $\Delta I_{A_{1g}^{*}}$ of anti-Stokes and Stokes signals from theory (top) and experiment (bottom) at $F = 18~\mu$J\,cm$^{-2}$. 
\textbf{c,} Fluence dependence of the anti-Stokes spectra at t = 2 ps. The $A_{\mathrm{1g}}^{\mathrm{*}}$ mode persists over all fluences, while $A_{\mathrm{1g}}$ emerging above $F_c$. 
\textbf{d,} Fluence-dependent time-resolved reflectivity (TRR) dynamics, dominated by $A_{\mathrm{1g}}^{\mathrm{*}}$ oscillations. Oscillations persist at all fluences, but become rapidly damped as $F$ approaches $F_c$. The dynamics are normalized to the initial transient reflectivity change, which increases with fluence, making the coherent $A_{\mathrm{1g}}^{\mathrm{*}}$ oscillations appear reduced, although their absolute amplitude increases and eventually saturates (Supplementary Note~11).
\textbf{e,} Anti-Stokes fits of $A_{\mathrm{1g}}^{\mathrm{*}}$ peak parameters versus fluence. Top: phonon frequency shows softening; bottom: amplitude saturates above $F_c$.
\textbf{f,} TRR analysis of $A_{\mathrm{1g}}^{\mathrm{*}}$: both oscillation frequency and amplitude saturate near $F_c$, consistent with Raman results. 
\textbf{g,} Theoretical model reproduces the saturation behavior of the $A_{\mathrm{1g}}^{\mathrm{*}}$ amplitude observed in the experiment.}
\label{fig:3}
\end{figure}
To track the fate of the CDW phonon across the critical threshold, we turn to anti-Stokes TR-Raman, which more directly probes the phonon occupation (Fig.~\ref{fig:3}). As the pump fluence approaches $F_c \!\approx\! 180~\mu$J\,cm$^{-2}$, the anti-Stokes signal of the CDW mode $A_{\mathrm{1g}}^{\mathrm{*}}$ increases and becomes comparable to its Stokes counterpart within $\sim$1~ps (Fig.~\ref{fig:3}a, blue box), demonstrating strongly populated non-equilibrium state. Above $F_c$, the anti-Stokes intensity of $A_{\mathrm{1g}}^{\mathrm{*}}$ saturates and shows no further increase, while the anti-Stokes response of the emergent normal-phase $A_{\mathrm{1g}}$ mode grows, signalling a crossover to a different dynamical regime.

\begin{figure}[ht!]
  \centering
  \includegraphics[width=0.72\textwidth]{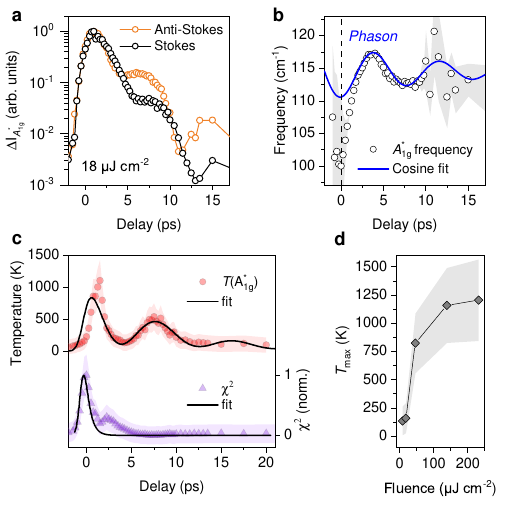} 
  \caption{\textbf{Comparison of Stokes and anti-Stokes dynamics, and extracted phonon temperature.}
  \textbf{a,} Stokes and anti-Stokes dynamics of $A_{\mathrm{1g}}^{\mathrm{*}}$  at $F = 18~\mu\mathrm{J\,cm}^{-2}$. The anti-Stokes signal closely tracks the coherent oscillations seen in the Stokes channel with enhanced relative amplitude, demonstrating the coupling of the locked phason to the CDW phonon.
  \textbf{b,} Time-dependent frequency of the $A_{\mathrm{1g}}^{\mathrm{*}}$ mode $\Omega_{A_{1g}^{*}}(t)$. The mode exhibits a pronounced initial softening, followed by a cosine oscillatory response of the same period as the locked phason.
  \textbf{c,} Extracted phonon temperature $T_{\mathrm{ph}}(t)$ and time-dependent Raman susceptibility $\chi(t)$ at $F = 46~\mu\mathrm{J\,cm}^{-2}$. The data reveal a monotonically decaying $\chi(t)$, indicative of a loss of CDW-related susceptibility, together with oscillatory dynamics of $T_{\mathrm{ph}}(t)$ that follow the low-frequency modulation of the $A_{\mathrm{1g}}^{\mathrm{*}}$ mode.
  \textbf{d,} Maximum phonon temperature $T_{\mathrm{ph}}^{\max}$ as a function of fluence, extracted from fits to the anti-Stokes dynamics. The apparent divergence and subsequent saturation of $T_{\mathrm{ph}}^{\max}$ above the critical fluence $F_c$ indicate that the CDW phonon enters an overheated regime.}
  \label{ED3}
\end{figure}

To obtain a quantitative estimate of the phonon population after photoexcitation, we evaluate the detailed anti-Stokes to Stokes ratio $I_\mathrm{AS}(t)/I_\mathrm{S}(t)$  and invert the Bose-Einstein relation (Supplementary Note~5, Eqs.~6,7) to extract the phonon occupation at frequency $\Omega$ and delay $t$, $n(\Omega,t)$, and the corresponding effective phonon temperature $T(t)$. Well below $F_c$, the maximum $T(t)$ remains moderate, around $100$-$200$\,K (Extended Data Fig.~\ref{ED3}). As $F$ approaches $F_c$, $T_{A_{1g}^{*}}$ rises steeply and exceeds $10^3$\,K, far above the equilibrium $T_\mathrm{CDW}$. Crucially, even at these extreme temperatures, a clear CDW phonon peak $A_{\mathrm{1g}}^{\mathrm{*}}$ remains in the Raman spectrum (Fig.~\ref{fig:3}a), demonstrating that the phonon is still present and that the PLD survives overheating of the CDW mode. The temporal profile of $T(t)$ shows a delayed rise and a slow oscillatory tail (Extended Data Fig.~\ref{ED3}). A fit with two exponential decay plus a damped cosine function (Supplementary Note~6, Eq.~8) yields a low-frequency mode near 0.13\,THz (Extended Data Fig.~\ref{ED3}). In contrast, the time-dependent Raman susceptibility decays exponentially on a $\sim$1.5\,ps timescale and shows no detectable modulation (Extended Data Fig.~\ref{ED3}). The coherent modulation of the $A_{\mathrm{1g}}^{\mathrm{*}}$ mode persists on the anti-Stokes side (Fig.~\ref{fig:3}b), and the effective theory closely reproduced its dynamics. The effect is more pronounced than on the Stokes side (Extended Data Fig.~\ref{ED3}), suggesting that this long-period mode interacts with a genuine phonon contribution rather than an electronic response. This mode appears with a cosine-like phase in the phonon sector (Frequency of anti-Stokes $A_{\mathrm{1g}}^{\mathrm{*}}$ and effective phonon temperature $T_{A_{1g}^{*}}$ shown in Extended Data Fig.~\ref{ED3}b and c), consistent with a hybrid excitonic-phononic character and a rapid displacive quench of the potential: photoexcitation changes the minimum set by the excitonic order on an ultrafast timescale, and the locked phason corresponds to coherent phase oscillations around this new minimum.

The fluence-delay map of the anti-Stokes signal in Fig.~\ref{fig:3}c shows that, while the normal-phase $A_{\mathrm{1g}}$ response strengthens dramatically above $F_c$, the CDW phonon $A_{\mathrm{1g}}^{\mathrm{*}}$  remains clearly visible at all fluences, demonstrating that the underlying PLD is not completely melted, but survives at least partially even in the high-fluence regime. Gaussian fits to the anti-Stokes lineshape reveal that, above $F_c$, the $A_{\mathrm{1g}}^{\mathrm{*}}$ frequency and amplitude approach plateau values (Fig.~\ref{fig:3}e): the mode neither softens further nor gains additional population. To independently probe the coherent dynamics, we performed TRR measurements across a broad fluence range and compared the results with anti-Stokes TR-Raman. The reduced oscillation visibility in Fig.~\ref{fig:3}d at high fluence stems from normalization to the large initial step and increased damping (Supplementary Note.~ 11). The reflectivity oscillations, matching the $A_{\mathrm{1g}}^{\mathrm{*}}$ frequency, show the same fluence dependence as the anti-Stokes Raman signal: amplitude saturation and frequency softening above $F_c$, with no evidence of phonon disappearance. By isolating the oscillatory component from the background decay and fitting the resulting signal, we extracted fluence-dependent amplitude and frequency trends (Fig.~\ref{fig:3}f), which closely mirror the frequency-domain results. 

Taken together, these observations reveal that the CDW mode does not melt away, but remains as a well-defined folded CDW phonon whose coherence is strongly suppressed once the critical regime is reached. Experimentally, this manifests as (i) a simultaneous saturation of the anti-Stokes $A_{\mathrm{1g}}^{\mathrm{*}}$ intensity and the phonon occupation $n(\Omega)$, such that $I_\mathrm{AS}(t) \approx I_\mathrm{S}(t)$ while the $A_{\mathrm{1g}}^{\mathrm{*}}$ peak itself remains clearly visible, and (ii) a saturation of the intensity of the CDW mode above $F_c$, verified by both TR-Raman and TRR: additional photoexcitation energy neither enhances the coherent oscillation amplitude nor further increases the phonon population. Instead, the $A_{\mathrm{1g}}^{\mathrm{*}}$ mode persists as a well-defined CDW phonon with essentially fixed spectral weight and frequency, while its phase-coherent motion is largely quenched and the PLD survives in an overheated, metastable configuration. 

This behaviour is captured by our effective theory (see Appendix, Effective theory), where the CDW phonon amplitude itself saturates as a function of pump strength (Fig.~\ref{fig:3}h), in line with the experimentally observed saturation of the TR-Raman anti-Stokes population (Fig.~\ref{fig:3}f). We emphasize that the effective model is designed to capture the symmetry structure and qualitative evolution of the coupled exciton-phonon potential, rather than to predict absolute frequencies or microscopic parameters, and therefore, its role is to elucidate the origin of the observed soft mode and the collapse of phase stiffness under increasing fluence. In the model, the amplitude phonon couples linearly to the excitonic order parameter $\Delta$, and a finite $\Delta$ value at low fluence stabilizes a Mexican-hat potential for the hybrid exciton-phonon order. The coupling to the phonon explicitly breaks the continuous phase symmetry of the excitonic condensate, giving rise to a finite-frequency pseudo-Goldstone excitation, the locked phason. As the fluence approaches the critical threshold $F_c$, the collapse of $\Delta$ gradually deforms the potential landscape and suppresses coherent phase oscillations, thereby removing the locked phason and reducing the restoring force that sustains the coupled dynamics. Experimentally, this corresponds to a saturation of both the anti-Stokes population and the coherent amplitude  (Fig.~\ref{fig:3}f,g), marking the onset of a regime dominated by incoherent phonon excitation.

\subsection*{Fingerprints of a Dynamically Trapped Non-Thermal State}

\begin{figure}[ht!]
    \centering
    \includegraphics[width=\linewidth]{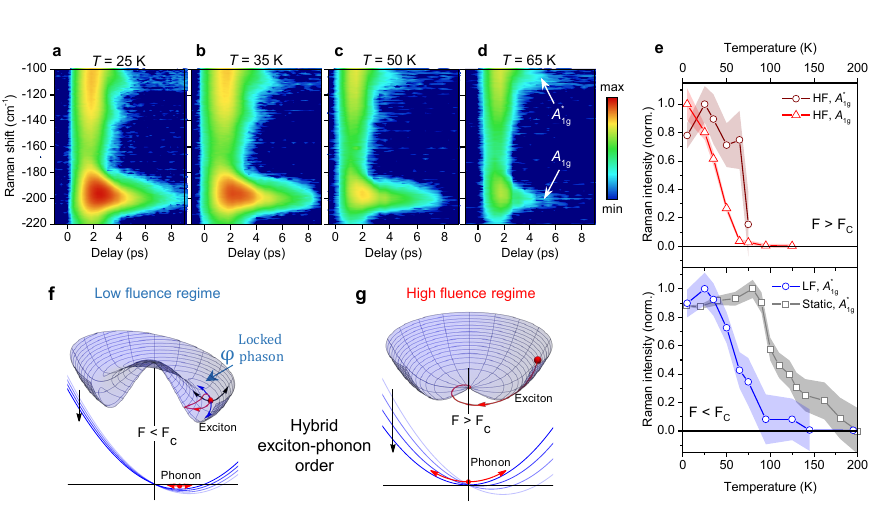}
    \caption{\textbf{The high-fluence regime constitutes a non-thermal, dynamically trapped phase.}
    \textbf{a-d,} Anti-Stokes TR-Raman maps $\Delta I$ at a fixed fluence of $F = 405~\mu\mathrm{J\,cm}^{-2}$ for base temperatures of 25, 35, 50, and 65~K. The photoinduced normal-phase $A_{\mathrm{1g}}$ mode ($\sim 200$~cm$^{-1}$, bottom arrow) weakens with increasing $T$ and vanishes by 65~K, whereas the folded CDW mode $A_{\mathrm{1g}}^{\mathrm{*}}$ ($\sim 116$~cm$^{-1}$, top arrow) persists.
    \textbf{e,} Temperature dependence of normalized Raman intensities. Top: In the high-fluence regime ($F>F_c$), both $A_{\mathrm{1g}}$ (red triangles) and $A_{\mathrm{1g}}^{\mathrm{*}}$ (red circles) collapse near 65~K. Bottom: In the low-fluence regime ($F<F_c$), the $A_{\mathrm{1g}}^{\mathrm{*}}$ intensity (blue circles) tracks the equilibrium static CDW order parameter (grey squares).
    \textbf{f, g,} Schematic of the coupled exciton-phonon potential. For simplicity, the excitonic potential is shown statically, and the red arrow indicates the trajectory of the order parameter after the quench. \textbf{f,} Low-fluence regime ($F < F_c$) where the tilted Mexican-hat potential remains intact, supporting coherent oscillations of locked phason. \textbf{g,} High-fluence regime ($F > F_c$) where the potential flattens, trapping the system in a metastable remnant PLD distinct from the thermal state.}
    \label{fig:4}
\end{figure}

Finally, we demonstrate that the high-fluence regime above $F_c$ constitutes a distinct, dynamically trapped non-equilibrium phase, one that cannot be accessed via thermal pathways. To probe this regime, we perform temperature-dependent TR-Raman at a fixed fluence well above $F_c$. The results reveal a strikingly non-thermal behavior: the intensity of the photoinduced $A_{\mathrm{1g}}$ mode, i.e., an excitation fingerprint of the normal phase, is strongest at low temperatures and systematically weakens as the base temperature approaches $T_\mathrm{CDW}$ (Fig.~\ref{fig:4}a-d). This trend is counterintuitive: in a thermally driven transition, one would expect the normal-mode population to grow with temperature, not decrease. Even at 65~K, the CDW mode remains comparable in strength to the $A_{\mathrm{1g}}$ (white arrows, Fig.~\ref{fig:4}d), highlighting their distinct dynamical origins.

This anomalous temperature dependence implies that a hidden, non-thermal parameter governs the evolution, most plausibly, the underlying electronic structure. This interpretation is consistent with the temperature dependence of the LERC.  As established by TR-ARPES, the electronic gap in 1\textit{T}-TiSe$_2$ remains partially open even at high fluence~\cite{huberRevealing2022,huberUltrafastCreationLightinduced2024,hedayatExcitonic2019,duanOptical2021,hedayat2021investigation}, supporting the persistence of a correlated phase. As the base temperature rises, this gap narrows, reducing the density of states available for phonon-mediated relaxation and potentially suppressing electron-phonon coupling due to the evolving lattice and band structure at the Brillouin zone boundary. In this picture, as the gap closes at higher temperatures, the relaxation channel weakens, explaining the suppression of photoinduced increase in $A_{\mathrm{1g}}$ intensity with increasing $T$. The system thereby enters a regime where a normal-mode response is strongest at the lowest temperature, signifying a trapped out-of-equilibrium phase stabilized by electronic structure constraints.

To confirm that the high-fluence state represents a distinct non-equilibrium phase, we contrast phonon evolution with the melting behavior observed at low fluence (Figure~\ref{fig:4}e). At low fluence, the $A_{\mathrm{1g}}^{\mathrm{*}}$ intensity persists up to $\sim$100\,K, tracking a non-thermal melting pathway that accelerates relative to the equilibrium transition (Supplementary Note~9). The dynamics change entirely under high fluence. Here, the $A_{\mathrm{1g}}^{\mathrm{*}}$ intensity deviates sharply from this trend, vanishing abruptly around 65\,K in stark contrast to the BCS-like suppression~\cite{chenCharge2015} and the observed evolution for an equilibrium order parameter. This anomalous behavior provides direct evidence for a non-thermal state and suggests that the excitonic component of the coupled order parameter has been annihilated. Meanwhile, the amplitude of the $A_{\mathrm{1g}}$ mode weakens gradually as the base temperature increases, diminishing to a constant, room-temperature-level background by $\sim$65\,K. This strong quenching is consistent with the highly nonlinear excitation mechanism of this mode (Figure~\ref{fig:2}f); as the system heats up, the progressive thermal closure of the residual CDW gap reduces the density of photoexcited carriers available for multiphonon relaxation. A simple laser-induced heating cannot account for the transition at 65\,K~\cite{mohr-vorobevaNonthermal2011} (Supplementary Note~10), reinforcing the non-thermal nature of this dynamically trapped state.

\section*{Discussion}

Our results demonstrate that the paradox of disentangling cooperative CDW order in 1\textit{T}-TiSe$_2$ under strong drive is not merely a delay of the lattice response, but the catastrophic collapse of a genuinely hybrid exciton-phonon order. By isolating a collective low-frequency mode, accessed through the CDW phonon and identified as a locked phason of the coupled order parameter, we obtain a direct spectroscopic signature of the shared potential landscape linking excitonic and structural degrees of freedom. This 0.13~THz mode emerges from the hybrid potential and acquires a finite frequency through breaking of the excitonic \( U(1) \) symmetry by the CDW phonon. We note that the observation of the soft locked phason reflects the phase dynamics of the hybrid order and does not by itself quantify the relative energetic contributions of excitonic versus lattice components.

Tracking the locked phason across the critical excitation threshold allows us to probe the transformation of the exciton-phonon potential. Our effective theory shows that this mode is stabilized by the excitonic order parameter \( \Delta \), and that its suppression marks the collapse of the coupled potential. This drives the saturation of the \( A_{1g}^{*} \) amplitude. Beyond this point, coupled excitonic-phononic dynamics vanishes: energy is transferred to incoherent phonon excitations, leaving a structurally ordered but excitonically depleted state.

The persistence of this structural distortion after the loss of excitonic order warrants further scrutiny. Recent ultrafast low-energy electron diffraction experiments have similarly reported a \textit{saturation} of PLD that survives at high fluence, consistent with our observations~\cite{kurtzNonthermal2024}. While that study attributes the remaining distortion to a dominant electron-phonon driven (Peierls-like) component of the equilibrium CDW, our detection of the soft locked phason challenges this interpretation. In our picture, the PLD survives not as a purely lattice-stabilized order, but as a dynamically arrested remnant. As the optical drive suppresses \( \Delta \), the hybrid potential flattens, removing the restoring force for the locked phason and leading to the saturation of the CDW phonon response (Fig.~\ref{fig:3}f,g). The lattice remains stranded in a non-thermal configuration due to inertial, anharmonic, and dissipative constraints that prevent structural recovery. This mechanism demonstrates how non-equilibrium drives can be used to dismantle complex quantum orders by dynamically reshaping the potential that binds their constituents.

This work reveals a new paradigm for quantum control: the dismantling of cooperative phases through nonlinear excitation. This mechanism offers a plausible connection to the light-induced dimensional crossover recently reported in this and other layered materials~\cite{chengLightinduced2022,duanOptical2021,domroseLightinduced2023,chengUltrafastFormationTopological2024}. In particular, the 3D-to-2D CDW crossover in 1\textit{T}-TiSe$_2$ has been linked to the breakup of bound electron-hole pairs~\cite{chengLightinduced2022}. The exciton-phonon catastrophe identified here provides a clear fluence-dependent pathway for annihilating the excitonic component of the order, potentially triggering the loss of interlayer coherence. These results suggest that direct manipulation of the low-frequency hybrid locked phason, such as via resonant excitation, could enable coherent control over the dimensionality of electronic order in layered quantum materials. Looking ahead, the strategy established here should be transferable to a broad class of correlated systems, using ultrafast optical drive as a practical design tool for non-equilibrium quantum phases.

\bibliography{new}

\begin{thebibliography}{}

\bibitem{Versteeg2018} R. B. Versteeg, J. Zhu, P. Padmanabhan, C. Boguschewski, R. German, M. Goedecke, P. Becker, and P. H. M. Van Loosdrecht, Struct. Dyn. \textbf{5}, 044301 (2018).
\bibitem{Bianco2015} R. Bianco, M. Calandra, and F. Mauri, Phys. Rev. B \textbf{92}, 094107 (2015).

\bibitem{Murakami2017} Y. Murakami, D. Golež, M. Eckstein, and P. Werner, Phys. Rev. Lett. \textbf{119}, 247601 (2017).
\bibitem{Ning2020} H. Ning, O. Mehio, M. Buchhold, T. Kurumaji, G. Refael, J. G. Checkelsky, and D. Hsieh, Phys. Rev. Lett. \textbf{125}, 267602 (2020).
\bibitem{Holt2001} M. Holt, P. Zschack, H. Hong, M. Y. Chou, and T.-C. Chiang, Phys. Rev. Lett. \textbf{86}, 3799 (2001).

\bibitem{Kresse1999} G. Kresse and D. Joubert, Phys. Rev. B \textbf{59}, 1758 (1999).
\bibitem{Hafner2008} J. Hafner, J. Comput. Chem. \textbf{29}, 2044 (2008).

\bibitem{Grimme2004} S. Grimme, J. Comput. Chem. \textbf{25}, 1463 (2004).
\bibitem{Duong2015} D. L. Duong, M. Burghard, and J. C. Schön, Phys. Rev. B \textbf{92}, 245131 (2015).
\bibitem{Togo2015} A. Togo and I. Tanaka, Scripta Mater. \textbf{108}, 1 (2015).
\bibitem{Monkhorst1976} H. J. Monkhorst and J. D. Pack, Phys. Rev. B \textbf{13}, 188 (1976).
\end{thebibliography}

\section*{Methods}

\subsection*{Time-resolved Raman spectroscopy (TR-Raman)}

Time-resolved coherent Raman measurements were performed in fs pump and ps probe configuration as detailed in \cite{Versteeg2018}. A 750\,nm $\sim$0.3~ps laser pulse was used as the pump to impulsively excite the sample, while a time-delayed 513\,nm $\sim$1.2~ps probe pulse was used to record the Stokes and anti-Stokes Raman spectra in parallel polarization geometry. The time resolution of the setup was approximately 1.2\,ps, defined by the cross-correlation between pump and probe pulses. A mechanical beam blocker cuts the scattering below 100\,cm$^{-1}$  to enable detection of the relevant low-energy phonon modes. The probe fluence was kept low to ensure a perturbative measurement, and the sample was held at cryogenic temperatures in a vacuum cryostat throughout all measurements.

\subsection*{Effective theory}

TiSe$_2$ undergoes a second-order transition at $T_c \approx 200~\mathrm{K}$ into a commensurate $2\times2\times2$ charge-density wave (CDW) with a weak trigonal lattice distortion. 
Experimental and theoretical studies point toward a cooperative origin involving both excitonic condensation and a periodic lattice distortion (PLD). 

To describe the ultrafast photoexcitation dynamics in 1\textit{T}-TiSe$_2$, we employ a minimal model consisting of two electronic bands (valence and conduction) and a single optical phonon mode. Electrons and phonons are itinerant, but their mutual interactions are local in real space. 
This approximation rests on three physical arguments:
(i) Coulomb screening renders the effective electron-electron interaction short-ranged,
(ii) the low-temperature CDW corresponds to an equal-weight superposition of the three symmetry-equivalent $M$-point phonons, so one effective phonon mode ($A_{\mathrm{1g}}^{\mathrm{*}}$) suffices (this is consistent with the literature, e.g., shown in DFT simulations \cite{Bianco2015}),
and (iii) the lattice distortion couples predominantly to on-site charge transfer between Ti and Se orbitals.\\

\noindent\textbf{1.~  Hamiltonian}\\

The total Hamiltonian
\begin{align}
H = H_\mathrm{el} + H_\mathrm{ph} + H_\mathrm{el-ph} + H_\mathrm{el-el}
\end{align}
comprises electronic, phononic, and interaction terms.  
Electrons in valence and conduction bands ($i=1,2$) have dispersions $\epsilon_i(\mathbf{k})$ in the Brillouin zone, while phonons correspond to a single local mode of frequency $\Omega$ at each site $n$:
\begin{align}
H_\mathrm{el} = \sum_{\mathbf{k},i} \epsilon_i(\mathbf{k}) 
\, \hat{c}^\dagger_{\mathbf{k},i} \hat{c}_{\mathbf{k},i}, 
\qquad 
H_\mathrm{ph} = \frac{1}{2M} \sum_n (\hat P_n^2 + M^2\Omega^2 \hat X_n^2),
\end{align}
where $\hat X_n$ and $\hat P_n$ are the displacement and conjugate momentum of site $n$. 

The precise form of these interactions is irrelevant; only their symmetry properties matter. The simple choices made here simplify the derivation of the equations of motion. Assuming short-range interactions, the electron-phonon and screened Coulomb terms are local and of Fröhlich, respectively, Hubbard form:
\begin{align}
H_\mathrm{el-ph} &= g \sum_{n,ij} \hat X_n 
\left( \hat{c}^\dagger_{n,i}\hat{c}_{n,j} + \mathrm{h.c.} \right),\\
H_\mathrm{el-el} &= V \sum_{n,ij}
\hat{c}^\dagger_{n,i}\hat{c}^\dagger_{n,j}
\hat{c}_{n,j}\hat{c}_{n,i}.
\end{align}
Here $g$ is the on-site electron-phonon coupling, and $V>0$ the screened electron-electron repulsion. 
The latter favors local electron-hole pairing between valence and conduction states, promoting excitonic condensation.  

The complex order parameter
\begin{align}
\Delta_n = V \sum_{ij} \langle \hat{c}^\dagger_{n,i} \hat{c}_{n,j} \rangle
\end{align}
quantifies the exciton condensate, which induces hybridization between the two bands. We assume that in the CDW phase, $\Delta_n$ and the lattice displacement $X_n$ share the same modulation wavevector: phonons connect the valence-band maximum at $\Gamma$ with the conduction-band minima at the symmetry-equivalent $M$ points, so a static distortion at momentum $\mathbf{q}_i=\Gamma\to M_i$ induces an excitonic gap with the same periodicity.

In thermal equilibrium, all $M$-points are symmetry equivalent. Here we assume that this holds also after photo-excitation, i.e., that the pump couples equally to all $M$ points, and hence only the uniform component of the coupled order is dynamic. We mention that this approach also excludes the effect of a collective, coherent rotation through different $M$-points, which might emerge from an anisotropic excitation of the material.   

We extend the time-dependent mean-field theory that was successfully used in the related material $\mathrm{Ta_{2}NiSe}_{5}$ \cite{Murakami2017,Ning2020} to the study of $\mathrm{TiSe}_{2}$:
we apply a mean-field decoupling to $H_\mathrm{el-ph}$ and $H_\mathrm{el-el}$ and assume homogeneous mean fields $\Delta_n \to  \Delta(\mathbf{q}_\ell)\equiv \Delta$ and $\langle \hat X_n\rangle\to \langle \hat X(\mathbf{q}_\ell)\rangle\equiv X$ for all components with  $\mathbf{q}_\ell$ being the vector pointing from $\Gamma$ to $M_\ell$. Moving to the reduced Brillouin zone, this yields 
\begin{align}
H^{\mathrm{MF}} =
\sum_{\mathbf{k},ij} 
\left[\epsilon_i(\mathbf{k})\delta_{ij}
-(\Delta + gX)\tau^x_{ij}\right]
\hat{c}^\dagger_{\mathbf{k},i}\hat{c}_{\mathbf{k},j}
+ g\,\mathrm{Re}(\Delta)\sum_n \hat X_n + H_\mathrm{ph},
\end{align}
where $\tau^x$ acts in the valence-conduction subspace.  
The combination $\Delta + gX$ forms an effective hybridization gap: a finite $\Delta$ induces $X$, and vice versa.\\

\noindent\textbf{2.~Time-Dependent Mean-Field Dynamics}\\

To model ultrafast pump-probe dynamics, we promote $\Delta(t)$ and $X(t)$ to time-dependent fields. 
Neglecting spatial variations, the system is described by a $(0+1)$D time-dependent Ginzburg-Landau theory (cf.\ Ref.~\cite{Ning2020}).
Implementing a canonical relaxation mechanism $\sim \kappa,\gamma$ driving the system back towards the stationary state yields the equations of motion
\begin{align}
\begin{split}
    \partial_t^2 X &=-\gamma \partial_t X-\Omega^2 X-2g\Omega\text{Re}\left(\Delta\right)\,\\
    i\partial_t \Delta &=(1+i \kappa(t))\left\{\left[-m(t)+ U|\Delta|^2\right]\Delta +2 g' X\right\}\,.
\end{split}
\end{align}
The phonon frequency $\Omega$ has been measured to high precision with x-ray scattering \cite{Holt2001}. The effective mass $m$ and anharmonicity $U$ are obtained from integrating over particle-hole fluctuations.
Here, the photon flux and the collective response of the electron fluctuations enter via the time-dependence in $m(t), \kappa(t)$
\begin{align}
\begin{split}
    m(t) &= m_0 - F e^{-t^2/\tau^2} - \eta(t), \\
    \kappa(t) &= \kappa_0 \bigl(1 + \eta(t)\bigr).
\end{split}
\end{align}

It includes the pump and a convolution of the pump with an exponential decay $\exp(-t/\tau_{th})\theta(t)$. The term $\eta(t)$ represents an effective Hartree term generated by the fluctuations generated during the quench
\begin{align}
    \eta(t)=F\eta_e e^{\frac{\tau ^2-4 t \tau_{th}}{4 \tau_{th}^2}}
   \left(\text{erf}\left(\frac{t}{\tau }-\frac{\tau }{2\tau_{th}}\right)+1\right).
\end{align}

Within this minimal model, the coupled dynamics of $\Delta(t)$ and $X(t)$ capture the coherent evolution of excitonic and phononic order observed in time-resolved Raman and reflectivity measurements. Without loss of generality, we set $m_0$ = 1. The other fixed parameters are $\Omega=\frac{37}{10}*2\pi$ (which translates to $\approx 3.5\, \text{THz}$ by setting the time unit to $1\,ps$), $\eta_e=1/40$, $U=1$, $\tau=1/10$, $g=-10$, $g'=-1/10$, $\gamma=1/2$, $\kappa=1/3$, $\tau_{th}=3$.\\

\noindent\textbf{3.~Collective Excitations and Locked Phason}\\

The excitonic-phononic order in 1\textit{T}-TiSe$_2$ breaks a discrete rotational symmetry rather than a continuous one. 
This symmetry breaking is governed by the electron-phonon coupling $g$, which is small compared to both $\Omega$ and the electron-electron interaction. 
Consequently, the phase (phason) mode of the exciton acquires a small mass term $\propto \sqrt{gg'}$, producing a weakly gapped collective excitation, the \emph{locked phason}.

Linearizing the equations of motion around equilibrium yields three collective modes:
(i) a high-frequency amplitude (\emph{Higgs}) mode, 
(ii) a hybridized phonon mode dominated by the PLD, and 
(iii) a low-frequency locked phason with $\omega_\mathrm{phason} \propto \sqrt{gg'} \ll \Omega$. 
This slow mode naturally explains the few-picosecond oscillations observed experimentally.

\subsection*{Density functional theory}

First-principles calculations were carried out using the Vienna Ab initio Simulation Package (VASP) \cite{Kresse1999, Hafner2008}. The structural coordinates for the commensurate $2\times2\times2$ CDW phase of 1\textit{T}-TiSe$_2$ (24 atoms) were taken from the GGA-vdW relaxed structure reported in Ref.~\cite{Bianco2015}, with one further relaxation step using the GGA-PBE exchange-correlation functional and projector augmented-wave (PAW) potentials. Van der Waals interactions were included using Grimme’s D2 correction \cite{Grimme2004}. Electronic smearing was applied with a broadening of 0.08\,eV (effective $T\sim 900$\,K), and results were checked down to $T\sim 100$\,K. Phonon properties were computed using Phonopy \cite{Togo2015} with a $2\times2\times2$ supercell (192 atoms) based on the CDW unit cell. A kinetic energy cutoff of 300\,eV and a $6\times6\times3$ Monkhorst-Pack $k$-point grid were used \cite{Monkhorst1976}.

\section*{Acknowledgements}
The authors gratefully acknowledge helpful discussions with Matteo Calandra, Giovanni Marini and Fulvio Parmigiani. The authors acknowledge financial support from the Deutsche Forschungsgemeinschaft (DFG, German Research Foundation) through CRC 1238 (Project No. 277146847, Control and Dynamics of Quantum Materials). Computational work was performed on the University of Bath’s High Performance Computing Facility, supported by the EU Horizon 2020 OCRE/GEANT project “Cloud funding for research”. G.C. acknowledges financial support from the European Union’s NextGenerationEU Programme via the I-PHOQS Infrastructure (IR0000016, ID D2B8D520, CUP B53C22001750006, ”Integrated infrastructure initiative in Photonic and Quantum Sciences”). C.J.S. and G.C. acknowledge support from the Horizon Europe EIC Pathfinder Open program under grant agreement No. 101130384 (QUONDENSATE).

\section*{Author Contributions}
H.H. conceived the project. O.A.A. performed the TR-Raman and D.C. conducted TR-reflectivity experiments. H.H. and P.v.L. supervised the study. J.L., N.B., M.B., S.D. developed the   theoretical model and performed the calculations. D.W. carried out the DFT calculations. C.J.S. provided the 1\textit{T}-TiSe$_2$ crystals. All authors- O.A.A., D.C., J.L., N.B., M.B., S.D., C.J.S., G.C., P.v.L., and H.H.- contributed to the discussions and interpretation of the results. H.H. and J.L. wrote the manuscript with contributions from O.A.A. and input from all authors.

\section*{Competing Interests}
The authors declare no competing interests.

\section*{Additional Information}
\textbf{Supplementary Information} The online version contains supplementary material available at https://doi.org/10.1038/XXXXX. \\
\textbf{Correspondence} and requests for materials should be addressed to P.v.L. or H.H.

\ExtendedDataFigures

\begin{figure}[p]
\centering
\includegraphics[width=0.72\textwidth]{GifEx2.pdf}
\caption{\textbf{Fluence-dependent behavior of the hybrid exciton-phonon mode.}
\textbf{a-c,} Residual coherent oscillations of the photoinduced $A_{1g}^*$ phonon dynamics after exponential background subtraction (Supplementary Note~6) for three representative fluences.
Extracted amplitude (\textbf{d}), damping time (\textbf{e}), and period (\textbf{f}) as a function of excitation fluence.
At low fluence, the oscillations are long-lived and periodic; with increasing fluence, the hybrid mode softens and damps, disappearing above $100~\mu\mathrm{J\,cm}^{-2}$.
This behavior indicates a fluence-driven collapse of the hybrid locked phason mode.}
\label{ED2}
\end{figure}

\begin{figure}[ht!]
  \centering
  \begin{subfigure}[b]{0.55\textwidth}
    \centering
    \includegraphics[width=\textwidth]{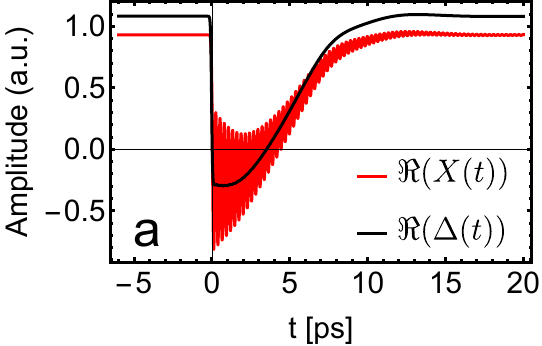}
  \end{subfigure}
  
  \vspace{1em} 
  
  \begin{subfigure}[b]{0.55\textwidth}
    \centering
    \includegraphics[width=\textwidth]{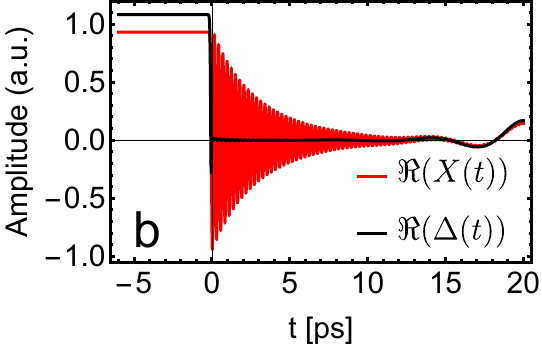}
  \end{subfigure}
  
  \caption{\textbf{Theoretical simulation of the exciton-phonon order.}
  Simulated time evolution of the real parts of the coupled order parameters following a sudden quench at time $t=0$. For low fluence ($F < F_c$) (\textbf{a}), the initial state is recovered quickly, following a phase of violent oscillations of the phononic order. For high fluence ($F > F_c$) (\textbf{b}), the excitonic order is lost, and the subsequent relaxation occurs on a much longer time scale. Lattice displacement is shown by $\mathfrak{R}(X(t))$  (red) and the excitonic order parameter by $\mathfrak{R}(\Delta(t))$ (black).}
  \label{ED4}
\end{figure}

\end{document}